# Room-temperature orbit-transfer torque enabling van der Waals magnetoresistive memories

Zhen-Cun Pan [a,1], Dong Li [b,1], Xing-Guo Ye [a,1], Zheng Chen [c,1], Zhao-Hui Chen [a], An-Qi Wang [a], Mingliang Tian [c,d,e], Guangjie Yao [a], Kaihui Liu [a], Zhi-Min Liao [a,f,*]

[a] *State Key Laboratory for Mesoscopic Physics and Frontiers Science Center for Nano-optoelectronics, School of Physics, Peking University, Beijing 100871, China*
[b] *Academy for Advanced Interdisciplinary Studies, Peking University, Beijing 100871, China*
[c] *Anhui Key Laboratory of Condensed Matter Physics at Extreme Conditions, High Magnetic Field Laboratory, Hefei Institutes of Physical Science, Chinese Academy of Sciences, Hefei 230031, China*
[d] *School of Physics and Optoelectronic Engineering, Anhui University, Hefei 230601, China*
[e] *Collaborative Innovation Center of Advanced Microstructures, Nanjing University, Nanjing 210093, China*
[f] *Hefei National Laboratory, Hefei 230088, China*

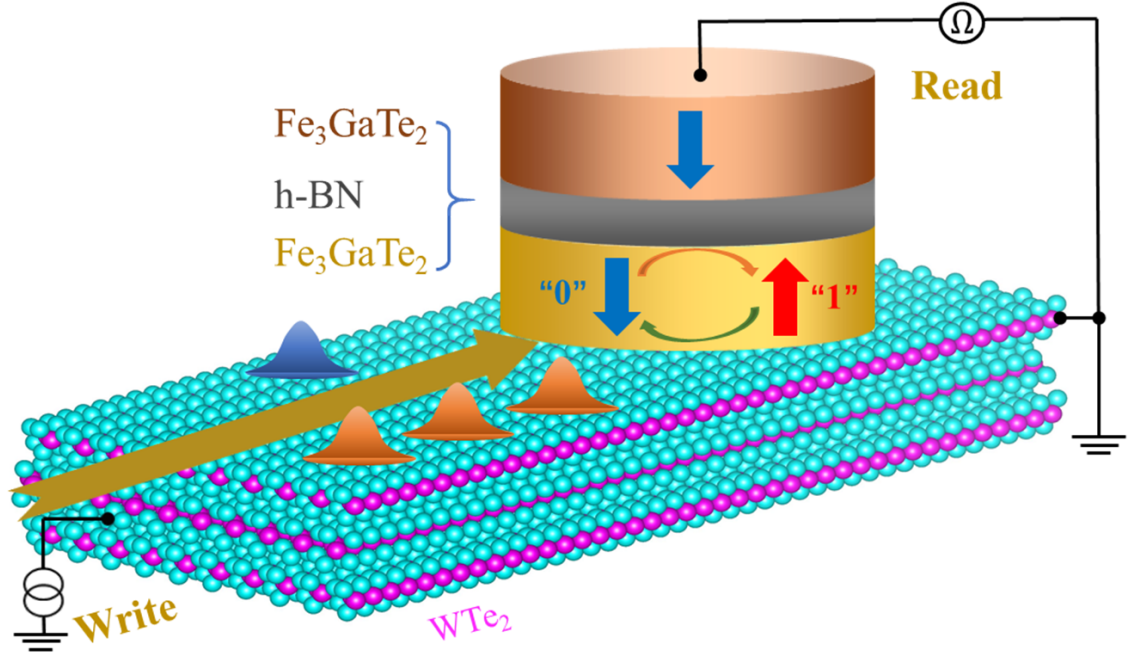

**Abstract:** The non-volatile magnetoresistive random access memory (MRAM) is believed to facilitate emerging applications, such as in-memory computing, neuromorphic computing and stochastic computing. Two-dimensional (2D) materials and their van der Waals heterostructures promote the development of MRAM technology, due to their atomically smooth interfaces and tunable physical properties. Here we report the all-2D magnetoresistive memories featuring all-electrical data reading and writing at room temperature based on $WTe_2/Fe_3GaTe_2/BN/Fe_3GaTe_2$ heterostructures. The data reading process relies on the tunnel magnetoresistance of $Fe_3GaTe_2/BN/Fe_3GaTe_2$. The data writing is achieved through current induced polarization of orbital magnetic moments in $WTe_2$, which exert torques on $Fe_3GaTe_2$, known as the orbit-transfer torque (OTT) effect. In contrast to the conventional reliance on spin moments in spin-transfer torque and spin–orbit torque, the OTT effect leverages the natural out-of-plane orbital moments, facilitating field-free perpendicular magnetization switching through interface currents. Our results indi-cate that the emerging OTT-MRAM is promising for low-power, high-performance memory applications.

## 1. Introduction

Magnetoresistive random access memory (MRAM), due to its high energy efficiency and fast data storage, has attracted tremendous attention in the era of big data [1–3]. In MRAM cells, data is stored as low and high resistance states in a magnetic tunnel junction (MTJ), achieved through parallel and antiparallel magnetization alignment of two ferromagnetic (FM) layers [4,5]. Writing data in MRAM cells relies on current-induced magnetization switching of the free FM layer in the MTJ [6–8]. With its non-volatile nature, MRAM shows promise for emerging applications like neural networks, in-memory computing, and stochastic computing [9]. The discovery of two-dimensional (2D) FM materials [10] with perpendicular magnetic anisotropy (PMA) enables compact device architectures, offering stable and reliable operation at the nanoscale to meet the demands of continuous MRAM miniaturization [3].

In conventional MRAM, energy-efficient data writing is accomplished using either spin-transfer torque (STT) [11,12] or spin–orbit torque (SOT) [13–18]. STT relies on spin-polarized current flowing through the device, while SOT switches magnetization by inducing interfacial spin polarizations between the FM and non-FM layers, thereby improving device durability and reliability [3,19]. These spin polarizations are generated through the spin Hall effect [7,8], Rashba-Edelstein effect [13], or topological surface states [20,21]. However, these effects typically produce in-plane spin polarization, which hinders magnetization switching with PMA [17]. Consequently, SOT-based PMA switching often requires an external magnetic field, although a few experiments have utilized SOT to achieve field-free PMA switching through structural asymmetry [22,23] or interfacial exchange bias [24].

Recently, a novel mechanism called orbit-transfer torque (OTT) has been proposed [25], which goes beyond spin-related effects like STT and SOT to achieve magnetization switching. OTT utilizes the current-induced orbital magnetization of Bloch electrons [26].

* Corresponding author.
  E-mail address: liaozm@pku.edu.cn (Z.-M. Liao).
[1] These authors contributed equally to this work.

In 2D systems, this orbital magnetization is naturally oriented out-of-plane, enabling the PMA switching in the adjacent FM layer without the need for an external magnetic field. Field-free magnetization switching induced by current has been achieved in WTe$_2$/Fe$_3$GeTe$_2$ heterostructures through the exploitation of the OTT effect [25], although some studies attribute similar experimental phenomena to unconventional SOT [27,28]. However, the operation temperature of these heterostructures remains below room temperature due to the limited Curie temperature of Fe$_3$GeTe$_2$, and practical realization of MRAM based on OTT is yet to be demonstrated.

Here, we report a room temperature operable all-2D magnetoresistive memories. It utilizes an OTT write mechanism and consists of a WTe$_2$/Fe$_3$GaTe$_2$/BN/Fe$_3$GaTe$_2$ heterostructure. Fe$_3$GaTe$_2$, a 2D FM metal with PMA and a Curie temperature above room temperature [29], forms an integral part of the MTJ, where the tunneling magnetoresistance (TMR) reaches 250% at low temperature and 14% at room temperature. By integrating a WTe$_2$ layer into the Fe$_3$GaTe$_2$-based MTJ, the magnetoresistive memory achieves field-free data writing and reading at room temperature.

## 2. Materials and methods

Fe$_3$GaTe$_2$ single crystals were grown using the Ga-Te flux method. High-purity Fe grains, Ga lumps, and Te chunks in a ratio of 1:1:2 were mixed and sealed in a vacuumed quartz tube. The tube was slowly heated to 950 °C in 12 h, and then kept at that temperature for another 12 h. After complete reaction, the tube was slowly cooled at a rate of 3 °C/h. Excess flux was removed through fast centrifugation at 780 °C. Shiny flake-like crystals were collected after natural cooling to room temperature.

High-quality bulk WTe$_2$ and h-BN crystals were purchased from HQ Graphene. Few-layer WTe$_2$, BN, and Fe$_3$GaTe$_2$ were obtained through standard mechanically exfoliated methods. Ti/Au electrodes (~10 nm thick) were patterned on individual SiO$_2$/Si substrates using standard e-beam lithography, metal deposition, and lift-off techniques. To achieve better contact, the electrodes were pre-cleaned using air plasma. Exfoliated BN flakes, few-layer Fe$_3$GaTe$_2$, and few-layer WTe$_2$ were sequentially picked up and then transferred onto the Ti/Au electrodes using a polymer-based dry transfer technique. The entire exfoliation and transfer processes were conducted in an argon-filled glove box with O$_2$ and H$_2$O content below 0.01 parts per million to prevent sample degradation.

Transport measurements were carried out in an Oxford cryostat with a variable temperature insert and a superconducting magnet. First- and third-harmonic voltage signals were collected by standard lock-in techniques (Stanford Research Systems Model SR830) with a frequency of $\omega$= 17.777 Hz, unless otherwise specified. A sequence of pulse-like current $I_p$ was applied through a Keithley 6221 current source. The pulse width was ~60 μs. After each $I_p$ was applied and then removed, the Hall resistance was measured by applying a small bias A.C. current (less than 5 μA).

Raman spectroscopy in parallel polarization was measured using 514 nm excitation wavelengths from a linearly polarized solid-state laser beam to determine the crystalline axis of WTe$_2$. The polarization of the excitation laser was controlled by a half-wave plate and a polarizer. The Raman scattered light with the same polarization as the excitation laser was collected.

## 3. Result and discussion

### 3.1. Field-free perpendicular magnetization switching

We initially demonstrate field-free magnetization switching via the OTT effect in a WTe$_2$/Fe$_3$GaTe$_2$ heterostructure (Fig. 1). OTT-driven magnetization switching is depicted in Fig. 1a and b. In low-symmetry 2D materials with nonzero Berry curvature dipole [30–32], such as T$_d$-phase WTe$_2$, an exotic orbital texture exists [26,33]. The polarization of orbital moment can be induced by a charge current along the crystalline $a$ axis in WTe$_2$. Conversely, a current along the $b$ axis (perpendicular to $a$ axis) of WTe$_2$ does not induce net orbital magnetization. When WTe$_2$ is adjacent to an FM layer, the orbital magnetization $\widehat{m}$ generates field-like torque $\tau^{FL} \sim M \times \widehat{m}$ and antidamping-like torque $\tau^{AD} \sim M \times (\widehat{m} \times M)$, where $M$ is the magnetization of the adjacent FM layer [19]. Unlike SOT, which typically involves in-plane spin polarization $\widehat{m}$, the OTT effect operates with out-of-plane $\widehat{m}$ due to dimensional constraints. For 2D FM materials with PMA, the existence of spin wave excitation can produce a perturbation of spin orientation deviating from the strictly perpendicular direction. Consequently, an out-of-plane antidamping-like torque can be exerted on the adjacent FM layer, enabling field-free PMA switching.

Following this principle, we fabricated WTe$_2$/Fe$_3$GaTe$_2$ heterostructures. Hall resistance measurements were conducted at various temperatures (Fig. S1 online). Fig. 1c shows the Hall resistance as a function of magnetic field at 300 K, revealing a significantly larger coercive field in the in-plane direction compared to the out-of-plane direction, which confirms the substantial PMA of Fe$_3$GaTe$_2$ at room temperature. To demonstrate OTT-driven magnetization switching, an $I_p$ is applied. To monitor the magnetization of Fe$_3$GaTe$_2$, the Hall resistance is measured using a small current ($i \ll I_P$) after the removal of $I_p$. The $I_p$ is initially applied approximately along the crystalline $a$ axis of WTe$_2$, as determined by angle-resolved transport in WTe$_2$ (Fig. S2 online). The Hall resistance as a function of $I_p$ along $a$ axis of WTe$_2$ at 300 K is shown in Fig. 1d. Intriguingly, a hysteresis loop with abrupt changes in Hall resistance is observed, indicating nonvolatile switching of magnetizations. Furthermore, the current-induced Hall resistance change precisely corresponds to the height of the $R_{xy} - \mu_0H$ loop, indicating full PMA switching. Importantly, no external magnetic field is applied for this $I_p$ induced PMA switching, consistent with the physical depiction of OTT in Fig. 1a and b.

Further, a series of current pulse along the $a$ axis in WTe$_2$ with an amplitude ±4.5 mA at 300 K were applied, as depicted in Fig. 1e. Robust magnetization switching was observed, with opposite-polarity current pulses resulting in termination magnetizations in opposite directions (up or down). Benefiting from the disc-like electrode structure, the current can be applied along various directions (Fig. S2a online). It is found that deterministic magnetization switching can be achieved at most angles due to the nonzero current component along $a$ axis (Fig. S3 online). Conversely, when applying current pulse along $b$ axis (perpendicular to $a$ axis) in WTe$_2$, no deterministic switching is observed, as illustrated in Fig. 1f. In this scenario, the $I_p$ only produced a termination state with nearly zero Hall resistance, which aligns with the multi-domain scenario induced by SOT [34].

### 3.2. Fe$_3$GaTe$_2$-based MTJ

To construct a magnetoresistive memory based on the WTe$_2$/Fe$_3$GaTe$_2$ heterostructure, the inclusion of a Fe$_3$GaTe$_2$-based MTJ is necessary. A typical MTJ device of a Fe$_3$GaTe$_2$/BN/Fe$_3$GaTe$_2$ heterostructure is schematized in the inset of Fig. 2a. The top and bottom Fe$_3$GaTe$_2$ layers were deliberately selected with different thicknesses to ensure distinct coercive fields. A thin BN layer (typically thinner than 3 nm) serves as the tunneling layer of the MTJ. The current–voltage ($I$–$V$) curve of the MTJ at 300 K is depicted in Fig. 2a, exhibiting typical nonlinear tunneling behavior.

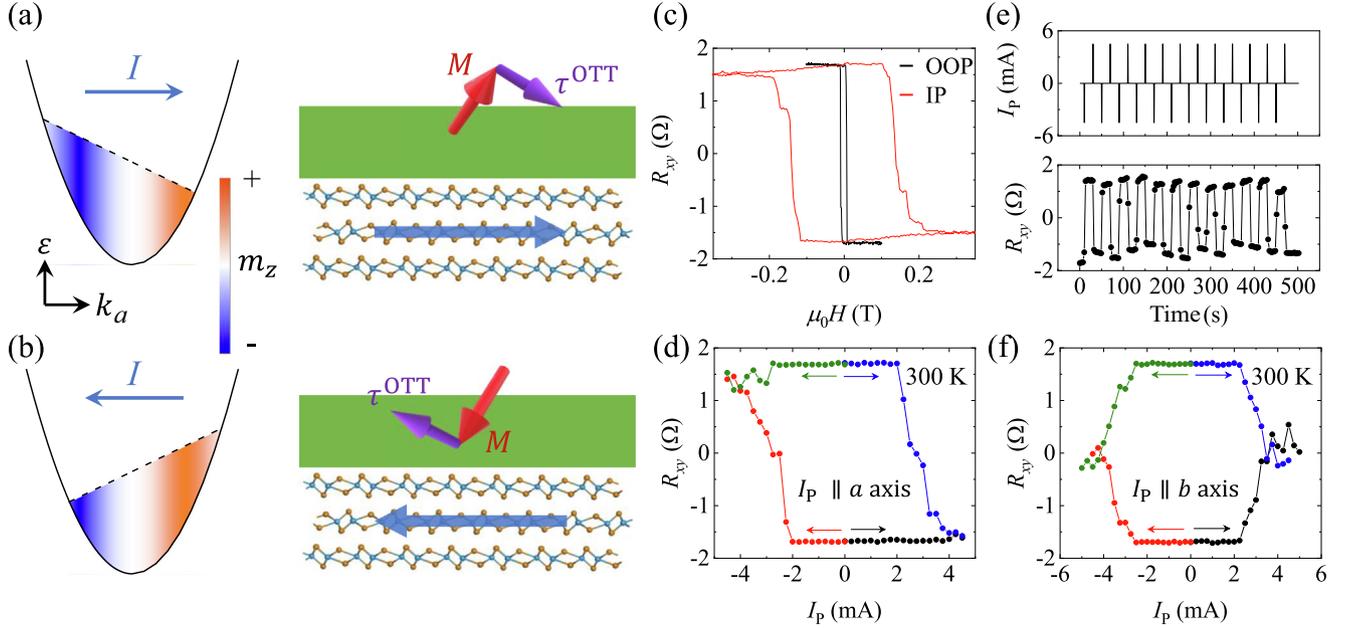

**Fig. 1.** (Color online) Current induced magnetization switching in a WTe$_2$/Fe$_3$GaTe$_2$ heterostructure at 300 K. (a), (b) Illustrations of OTT driven magnetization switching: current-induced magnetization of orbital magnetic moments (left column) and torque exerted by orbital magnetic moment in WTe$_2$ layer making PMA switching in Fe$_3$GaTe$_2$ layer (right column). (c) Hall resistance as a function of magnetic fields in out-of-plane (OOP) and in-plane (IP) directions. (d) Hall resistance as a function of pulse current ($I_p$) along crystalline $a$ axis of WTe$_2$. (e) Upper: applied series of current pulses with amplitude ±4.5 mA. Lower: corresponding changes in Hall resistance. (f) Hall resistance as a function of $I_p$ along crystalline $b$ axis of WTe$_2$.

By applying a small bias of 10 mV, the resistance of the MTJ was measured as a function of magnetic field at 300 K, as presented in Fig. 2b. Notably, the MTJ demonstrates distinct resistance states that correspond to the antiparallel and parallel magnetization alignment of the top and bottom Fe$_3$GaTe$_2$ layers, respectively. The MR is calculated using $MR = 100\% \times \frac{R(B)-R(0)}{R(0)}$, where $R(B)$ and $R(0)$ are resistances with and without magnetic field, respectively. Fig. 2b displays multiple resistance steps, which could stem from the existence of multiple magnetic domains in Fe$_3$GaTe$_2$, leading to various states of magnetization as sweeping the magnetic field.

The MTJ was investigated at various temperatures, and the measured MR are presented in Fig. 2c. Accordingly, the TMR ratio of the antiparallel magnetization state of the MTJ is extracted, and its temperature dependence is shown in Fig. 2d. At 10 K, the TMR reaches a remarkable value of 122% but gradually decreases with increasing temperature. The TMR remains 14% at room temperature. In the MTJ, the TMR relies on the spin polarization of the FM layers, governed by $TMR = \frac{2P^2}{1-P^2}$, where $P$ represents the spin polarization of the tunneling density of states at the Fermi level and is defined as $P = \frac{\mathcal{N}_{F\uparrow}-\mathcal{N}_{F\downarrow}}{\mathcal{N}_{F\uparrow}+\mathcal{N}_{F\downarrow}}$ (Refs. [35–37]). At 10 K, the TMR of ~122% corresponds to a majority spin percentage of 81%. The spin polarization $P$ is also plotted as a function of temperature in Fig. 2d, following $P = P_0 (1 - T/T_c)^\alpha$, where $P_0$ denotes the spin polarization at 0 K, $T_c$ is the critical temperature of the MTJ, and $\alpha$ is a material-dependent coefficient [35]. The fitting results in Fig. 2d yield a critical temperature of $T_c \sim 343$ K, higher than room temperature. Another Fe$_3$GaTe$_2$ based MTJ (Fig. S4 online) exhibits similar results, with a TMR of ~250% at 4 K and ~6% at room temperature and a $T_c$ of ~400 K.

### 3.3. Van der Waals magnetoresistive memories

The all-2D magnetoresistive memory is achieved by stacking the WTe$_2$ layer with Fe$_3$GaTe$_2$-based MTJ, as illustrated in Fig. 3a. The optical image in Fig. 3b shows a typical device with the follow-ing stacked structures from bottom to top: WTe$_2$, bottom Fe$_3$GaTe$_2$, thin h-BN, top Fe$_3$GaTe$_2$ and capping BN layers with thickness are 3.5, 6.8, 1.6, 15 and 30 nm, respectively. The crystalline axis of WTe$_2$ is identified by polarized Raman spectroscopy. The orientation dependence of the Raman peak at a Raman shift of ~210 cm$^{-1}$ is shown in Fig. 3c.

The TMR of the device measured at 300 K is shown in Fig. 3d. It is observed that the TMR is ~10% at a bias voltage of 5 mV and remains ~5% at 100 mV. Fig. 3e demonstrates the non-volatile information writing achieved by applying a writing current pulse ($I_W$) at 300 K. The reading of information, encoded as "0" and "1", is accomplished by measuring the resistance of the MTJ, corresponding to the low and high resistance states, respectively. By applying an $I_W$ with nonzero component along the $a$-axis in WTe$_2$, the produced OTT switches the magnetization of the adjacent (bottom) Fe$_3$GaTe$_2$ layer, while maintaining the magnetic state of the top Fe$_3$GaTe$_2$ layer unchanged. As a result, the OTT effect leads to the parallel and antiparallel alignment between the two Fe$_3$GaTe$_2$ layers, enabling the switching between "0" and "1" (the low and high resistance states). Furthermore, continuous write operations are demonstrated by applying a sequence of current pulses with an amplitude of ±1.8 mA. After each pulse is removed, the tunneling resistance is measured. Robust information writing and reading at room temperature are achieved, as illustrated in Fig. 3f. It is worth noting that such information writing through OTT is robust against in-plane magnetic field (Fig. S5 online). Thermal effect is ruled out as the driving factor behind magnetization switching in the WTe$_2$/Fe$_3$GaTe$_2$ memory device (Fig. S6 online).

### 3.4. Discussions on the mechanism of orbit-transfer torque

Generally, the magnetization of FM layer can be switched by the adjacent current-induced magnetic moment polarization $\sigma$. There exist two kinds of torques, i.e., field-like torque and antidamping-like torque, following $\boldsymbol{T}_{FL} = \tau_{FL}\boldsymbol{M} \times \boldsymbol{\sigma}$ and

Z.-C. Pan et al.

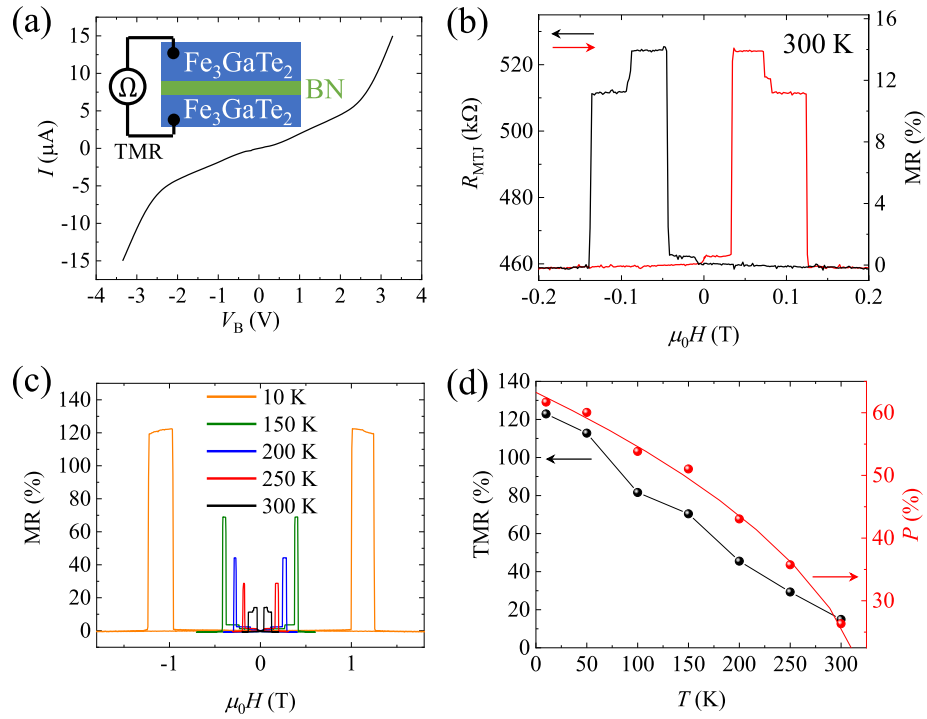

**Fig. 2.** (Color online) Fe$_3$GaTe$_2$-based MTJ. (a) *I–V* characteristics at 300 K, with the inset illustrating the MTJ configuration. (b) Tunneling resistance and the corresponding MR as a function of magnetic field at 300 K. (c) MR at various temperatures. (d) Temperature dependence of TMR ratios (black) of the antiparallel magnetization state of the MTJ and spin polarization ratio (red).

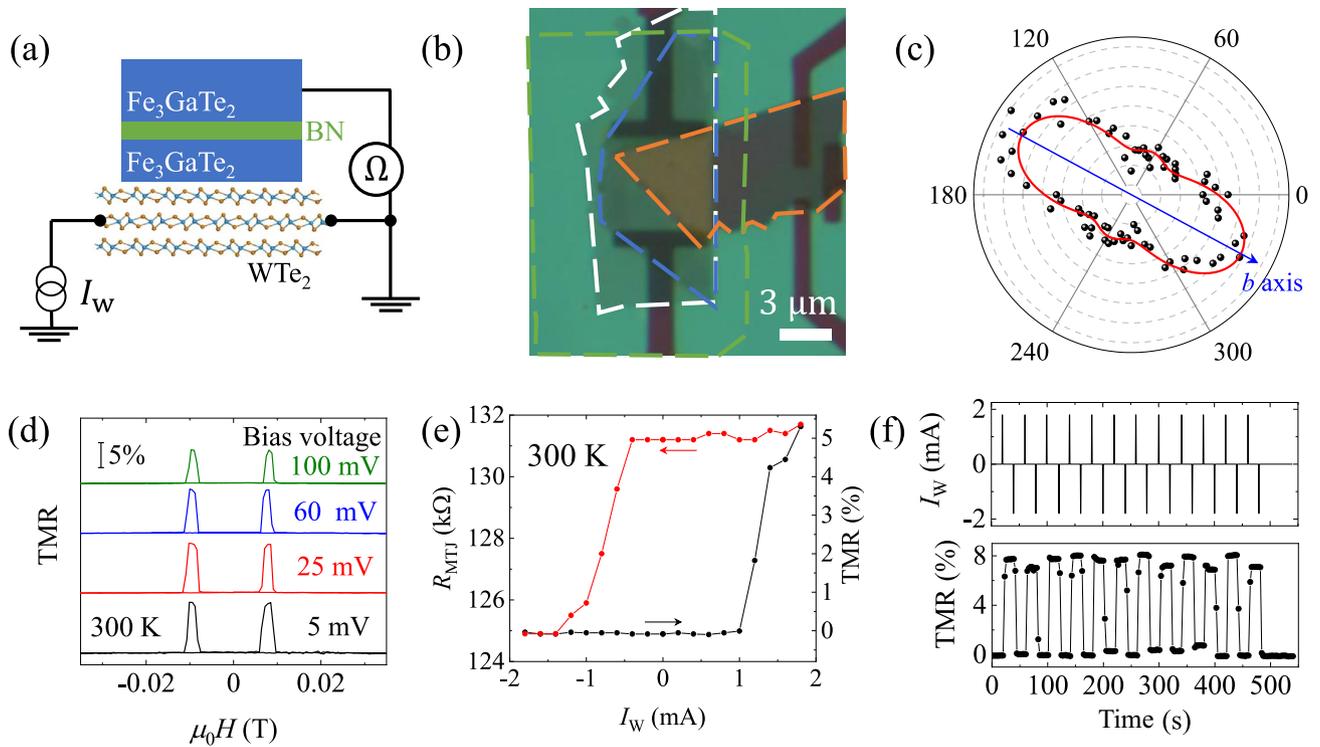

**Fig. 3.** (Color online) OTT-based magnetoresistive memory at room temperature. (a) Schematic of the WTe$_2$/Fe$_3$GaTe$_2$/BN/Fe$_3$GaTe$_2$ heterostructure for OTT-MRAM cell. (b) Optical image displaying a typical device with the heterostructure composed of WTe$_2$, bottom Fe$_3$GaTe$_2$, h-BN, and top Fe$_3$GaTe$_2$ layers marked by white, blue, green, orange dashed borders, respectively. (c) Polarized Raman spectroscopy, identifying the crystalline *b*-axis of WTe$_2$. (d) TMR at 300 K plotted as a function of magnetic field under various bias voltages. (e) The switching between high and low resistance states of the tunneling resistance by applying writing current ($I_W$) pulses at 300 K. (f) Upper: application of a series of current pulses with an amplitude of ±1.8 mA at 300 K. Lower: corresponding changes in TMR.

$T_{AD} = \tau_{AD} M \times (\sigma \times M)$, respectively, where $M$ is the magnetization of the FM layer. Usually, the magnetization switching is driven by the antidamping-like torque, which tends to force $M$ to be parallel to $\sigma$. Therefore, to exploiting in-plane $\sigma$ for perpendicular magnetization switching, an external magnetic field or additional asymmetric design is needed. In this case, the critical value of the torque is $\tau_{AD}^{\parallel} = \gamma H_{an}/2$, where $\gamma$ is the gyromagnetic ratio and $H_{an}$ is the perpendicular anisotropy field [8]. In most of the cases, the spin–orbit coupling generates in-plane $\sigma$, which is incompatible with perpendicular magnetization switching.

Interestingly, if an out-of-plane $\sigma$ is generated, field-free magnetization switching can be realized. In this case, the out-of-plane antidamping-like torque only needs a critical value around $\tau_{AD}^{\perp} = \gamma \alpha H_{an}$, where $\alpha$ is the Gilbert damping parameter. Because $\alpha$ is typically on the order of 0.01, the out-of-plane antidamping-like torque can realize energy-efficient magnetization switching with much lower critical current density.

The orbital magnetic moment, originated from self-rotating Bloch electrons, points along out-of-plane direction in 2D materials due to the dimensional constraint. Therefore, out-of-plane $\sigma$ can be generated by orbital polarization. It is reported that 2D materials with nonzero Berry curvature dipole can result in current-induced orbital polarization, following $\sigma \propto D \cdot E$, where $D$ is the Berry curvature dipole and $E$ is the applied electric field [32]. Note that few-layer $WTe_2$ holds nonzero Berry curvature dipole along $a$ axis on the surface. By applying current with nonzero component along $a$ axis of $WTe_2$, orbital polarization can be generated. Therefore, out-of-plane antidamping-like torque, that is, the OTT, exerts on the adjacent FM layer, which induces field-free perpendicular magnetization switching with high energy efficiency.

The observed magnetization switching in this work is quite consistent with the mechanism of OTT, due to the following reasons. First, the perpendicular magnetization switching is achieved without the assistant of external magnetic field. This is inconsistent with conventional SOT, where an external magnetic field or structure asymmetry is needed. Nevertheless, the current-driven orbital polarization is along out-of-plane direction, which can naturally lead to field-free perpendicular magnetization switching. Second, the field-free perpendicular magnetization switching is achieved only when existing nonzero current component along $a$ axis. When current is along $b$ axis, that is, perpendicular to $a$ axis, no deterministic switching is observed (Fig. 1f). This is well consistent with the mechanism of OTT, which is produced in the case of the existence of parallel Berry curvature dipole and electric field. Because the Berry curvature dipole in $WTe_2$ is along $a$ axis, only current component along $a$ axis can lead to OTT effect. Third, such magnetization switching is unaffected by applying in-plane magnetic field, which is consistent the mechanism of OTT but inconsistent with SOT. An out-of-plane spin polarization can be easily affected by an in-plane magnetic field, while on the contrary, the OTT with orbital polarization is robust against in-plane magnetic field due to dimensional constraint. We demonstrate the OTT effect under an in-plane magnetic field in a $WTe_2/Fe_3GeTe_2$ device (Fig. S5 online), where the in-plane magnetic field exerts minimal influence on the perpendicular magnetization switching.

Beyond $WTe_2$, alternative 2D materials possessing Berry curvature dipoles can also find application in OTT-based devices. Examples include $MoTe_2$, twisted bilayer graphene, strained $WSe_2$, among others. Typically, the magnitude of the Berry curvature is inversely related to the bandgap. As a result, materials with small bandgaps, like various topological Weyl semimetals with symmetry breaking, are anticipated to exhibit pronounced Berry curvature dipoles, making them viable candidates for applications in OTT-MRAM.

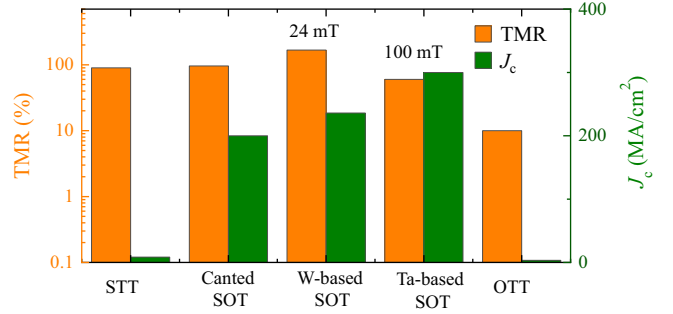

**Fig. 4.** (Color online) Comparative Analysis of STT-, SOT- and OTT-based data writing. Comparison of TMR ratio (orange) and critical switching current density ($J_c$) (green) at 300 K across different MTJ and magnetic switching technologies. Data collected from Ref. [40] for STT combined CoFeB/MgO/CoFeB, Ref. [41] for canted SOT based W/CoFeB/MgO/CoFeB, Ref. [40] for SOT-based W/CoFeB/MgO/CoFeB, and Ref. [42] for SOT-based Ta/CoFeB/MgO/CoFeB and our work for OTT.

## 4. Conclusion and outlook

The development of 2D materials has significantly enhanced the MRAM technology, thanks to their atomically smooth interfaces, tunable properties through proximity effects, and controllable crystal symmetry [3,10,38]. Previous studies have reported that 2D material-based MTJs exhibit comparable TMR values to conventional technologies. For instance, bilayer $CrI_3$ (Ref. [39]) and $Fe_3GeTe_2$-based MTJs [35] have TMR ~310% and ~160%, respectively. However, their performance at room temperature has been limited due to the constraint of Curie temperature. The discovery of the 2D $Fe_3GaTe_2$, which exhibits above-room-temperature ferromagnetism, has been beneficial for all-2D MTJs working at room temperature. In this work, we have observed significant TMR of ~14% at room temperature in $Fe_3GaTe_2$-based MTJs.

By using the engineered crystal symmetry of 2D materials, there are increased opportunities to achieve energy-efficient MRAM devices. The presented OTT effect enables field-free deterministic PMA switching by utilizing the unique orbital texture of $WTe_2$, which is facilitated by its low crystal symmetry. Fig. 4 provides a comparison between OTT and typical STT and SOT based MTJs. They exhibit similar TMR values at room temperature. Notably, the OTT-MTJ demonstrates the relatively low critical switching current density ($J_c$) required for information writing, highlighting its high energy-efficiency. The $J_c$ of OTT, ~3 × $10^6$ A/cm$^2$, is typically two orders of magnitude smaller than that of SOT [40–42]. The energy consumption for the data writing in the present device is about 1 fJ/nm$^2$ using 1.8 mA/60 μs pulses, which has ample room to be decreased through the utilization of shorter pulse durations, reducing device dimensions, and design optimization. The room-temperature dissipation power density of the OTT memory is ~1.5 × $10^{15}$ W/m$^3$, which is comparable to the heterostructures of $WTe_2$ and ferromagnet $Fe_3GeTe_2$ (Ref. [43]). The effective integration of 2D materials with distinct properties, as demonstrated in the case of OTT devices, is expected to drive the rapid development of ultrafast, energy-efficient, and non-volatile memories.

## Conflict of interest

The authors declare that they have no conflict of interest.

## Acknowledgments

This work was supported by the National Natural Science Foundation of China (61825401 and 91964201), and the Innovation Program for Quantum Science and Technology (2021ZD0302403).

Z.-C. Pan et al.

**Author contributions**

Zhi-Min Liao conceived and supervised the project. Zheng Chen and Mingliang Tian synthesized the $Fe_3GaTe_2$ crystals. Zhen-Cun Pan and Dong Li fabricated the devices and performed the transport measurements. Guangjie Yao and Kaihui Liu performed the Raman characterizations. Zhi-Min Liao, Xing-Guo Ye, Zhen-Cun Pan, Dong Li, Zhao-Hui Chen, and An-Qi Wang analyzed the data. Zhi-Min Liao and Xing-Guo Ye wrote the manuscript with discussion and input of all authors.

**Appendix A. Supplementary materials**

Supplementary materials to this article can be found online at https://doi.org/10.1016/j.scib.2023.10.008.

Supplementary materials for:

# Room-temperature orbit-transfer torque enabling van der Waals magnetoresistive memories

Zhen-Cun Pan[1,+], Dong Li[2,+], Xing-Guo Ye[1,+], Zheng Chen[3,+], Zhao-Hui Chen[1], An-Qi Wang[1], Mingliang Tian[3,4,5], Guangjie Yao[1], Kaihui Liu[1], Zhi-Min Liao[1,6*]

[1]State Key Laboratory for Mesoscopic Physics and Frontiers Science Center for Nano-optoelectronics, School of Physics, Peking University, Beijing 100871, China.

[2]Academy for Advanced Interdisciplinary Studies, Peking University, Beijing 100871, China.

[3]Anhui Key Laboratory of Condensed Matter Physics at Extreme Conditions, High Magnetic Field Laboratory, Hefei Institutes of Physical Science, Chinese Academy of Sciences, Hefei 230031, China.

[4]School of Physics and Materials Sciences, Anhui University, Hefei 230601, China.

[5]Collaborative Innovation Center of Advanced Microstructures, Nanjing University, Nanjing 210093, China.

[6]Hefei National Laboratory, Hefei 230088, China.

+ These authors contributed equally.

* Email: liaozm@pku.edu.cn

**Supplemental Note 1: Additional transport data of the WTe$_2$/Fe$_3$GaTe$_2$ heterostructure.**

The WTe$_2$/Fe$_3$GaTe$_2$ heterostructure was fabricated to demonstrate the field-free perpendicular magnetization switching at room temperature, as shown in Fig. 1 of main text. The Hall resistance as a function of out-of-plane magnetic field at various temperature is shown in **Fig. S1a**. The sharp hysteresis loop confirms the large perpendicular magnetic anisotropy in Fe$_3$GaTe$_2$. The anomalous Hall resistance shows nonmonotonic dependence on temperature (**Fig. S1b**). **Figure S1c** shows the temperature dependence of the coercive field, which is weakened by raising up temperature.

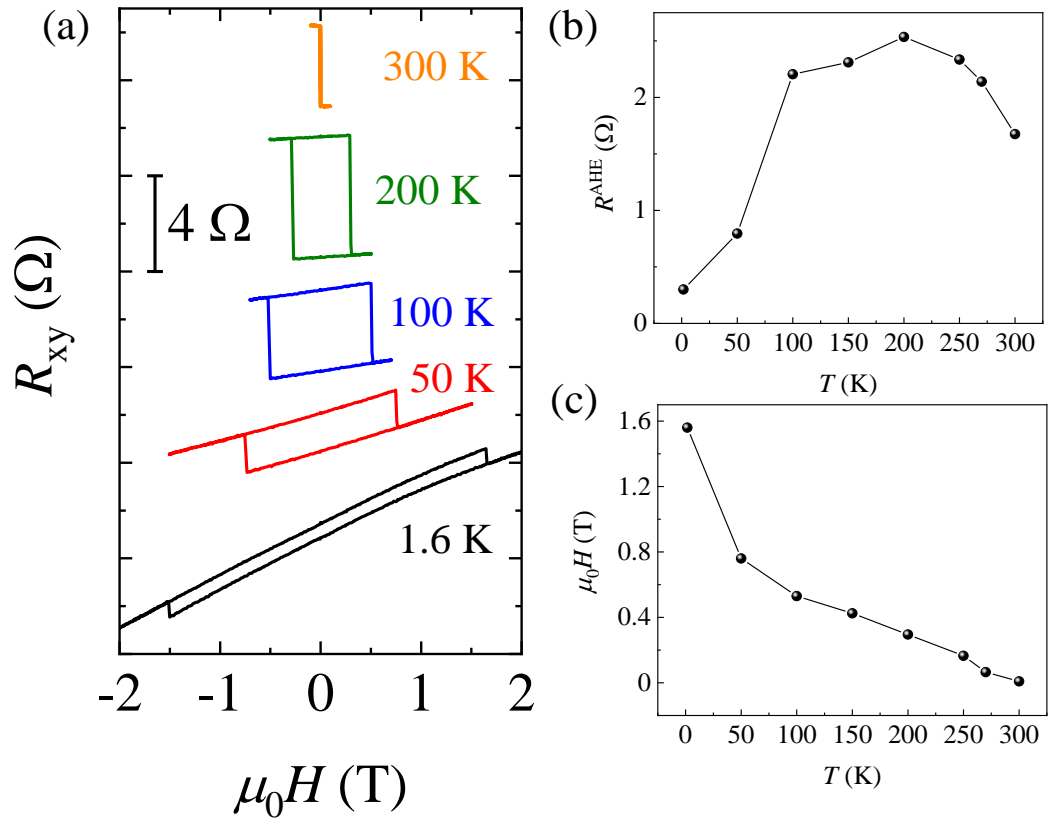

**Fig. S1. Magnetic properties of WTe$_2$/Fe$_3$GaTe$_2$ heterostructure. a,** The Hall resistance as a function of magnetic field at various temperatures. **b,** The anomalous

Hall resistance, defined as the half of the R-H loop height, as a function of temperature.

**c,** The coercive field as a function of temperature.

The angle-resolved transport in the heterostructure was investigated using the circular disc electrode structure shown in **Fig. S2a**. The angle $\theta$ is defined as the relative angle between the current direction and a baseline of the electrode pair, which is situated close to the crystalline *a*-axis of WTe$_2$. The longitudinal resistance $R_\parallel$ is plotted as a function of $\theta$ in **Fig. S2b**. It is fitted by the equation $R_\parallel(\theta) = R_a\cos^2(\theta - \theta_0) + R_b\sin^2(\theta - \theta_0)$, where $R_a$, $R_b$ is the resistance along the *a*- axis and *b*-axis, respectively. Here, $\theta_0$ corresponds to the angle misalignment between $\theta = 0°$ and the crystalline *a*-axis, and $\theta_0$ is ~3° for this device. Notably, the third-order nonlinear Hall effect, arising from the Berry connection polarizability, is observed (**Fig. S2c**). The angle dependence of the third-order Hall effect is given by

$$(V_H^{3\omega})/(V^\omega)^3 \propto \frac{\cos(\theta-\theta_0)\sin(\theta-\theta_0)[(\chi_{22}r^4-3\chi_{12}r^2)\cos^2(\theta-\theta_0)+(3\chi_{21}r^2-\chi_{11})\sin^2(\theta-\theta_0)]}{(\sin^2(\theta-\theta_0)+r\cos^2(\theta-\theta_0))^3},$$

where *r* is the resistance anisotropy, $\chi_{ij}$ are elements of the third-order susceptibility tensor. The fitting results in **Fig. S2d** also give rise to a misalignment $\theta_0 \sim 3°$.

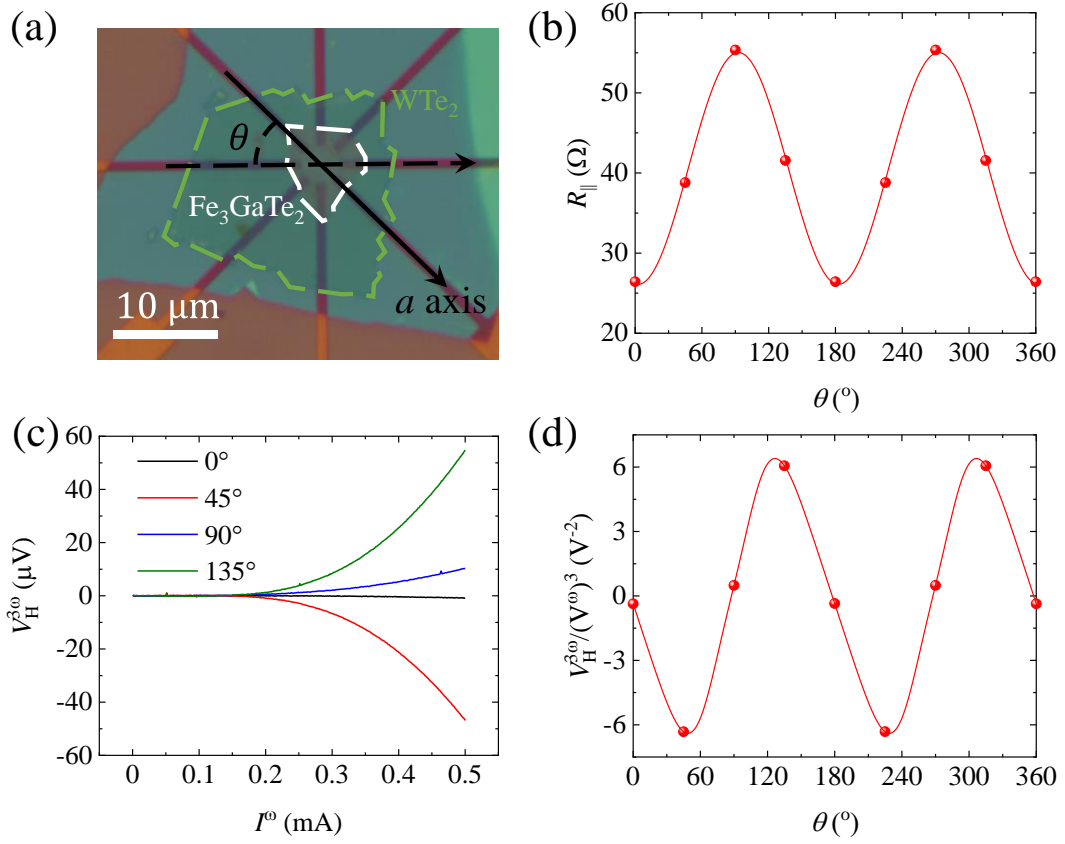

**Fig. S2. Anisotropic transport in WTe₂/Fe3GaTe2 heterostructure at 1.6 K. a,** Optical image of the device illustrating the angle $\theta$, defined as the angle between the applied current and the crystalline *a*-axis of WTe₂. **b,** Longitudinal resistance $R_\parallel$ as a function of $\theta$. **c,** Third-order Hall effect observed at various $\theta$. **d,** Variation of $\frac{V_H^{3\omega}}{(V^\omega)^3}$ with $\theta$.

Besides the *a*-axis and *b*-axis orientations, the pulse current for magnetic switching is also applied along the directions of $\theta = 45°$ and $\theta = 135°$. **Figure S3** demonstrates that deterministic magnetization switching is achieved at both angles. The relationship between the critical switching current ($I_c$) and crystal direction is shown below in **Table S1**. At $\theta = 90°$ (*b* axis), no deterministic switching is observed, where $I_c$ is ill-defined. In other angles, the $I_c$ shows a complicated dependence on $\theta$. This can be understood by the synergistic effect between OTT and thermal effect. When rotating $\theta$ away from 0°, the OTT is weakened due to the decrease of current component along $a$ axis. However, the thermal effect is enhanced when rotating $\theta$ away from 0° due to the resistivity anisotropy of WTe$_2$, making the magnetization switching easier. Therefore, the $I_c$ shows a minimum value at the intermediate angles.

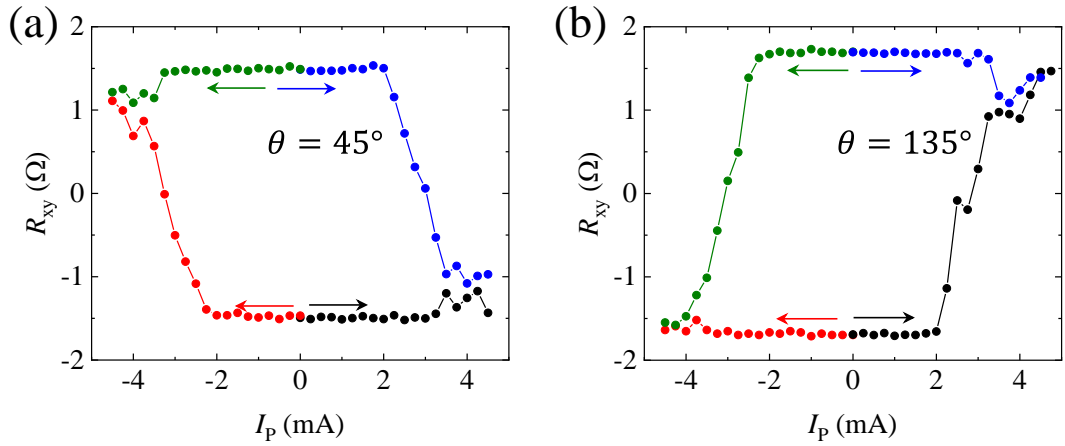

**Fig. S3. Field-free perpendicular magnetization switching using current applied off the crystalline axis. a,** and **b,** show the anomalous Hall resistance as a function of current pulse at 300 K, with current applied at $\theta = 45°$ and $\theta = 135°$, respectively.

**Table S1. The critical switching current versus current direction in WTe$_2$/Fe$_3$GaTe$_2$ heterostructure at 300 K.**

| $\theta$ (°) | $I_c$ (mA) |
|---|---|
| 0 | 4.54 |
| 45 | 4.52 |
| 90 | NA |
| 135 | 4.58 |

# Supplemental Note 2: Reproducible results from another Fe$_3$GaTe$_2$-based magnetic tunnel junction.

Additional Fe$_3$GaTe$_2$/BN/Fe$_3$GaTe$_2$ magnetic tunnel junctions (MTJs) were fabricated, yielding consistent results. **Figure S4a** demonstrates the MTJ's tunneling magnetoresistance (TMR) values, with ~250% at 4 K and ~6% at room temperature. The temperature dependence of TMR for the high resistance states is presented in **Fig. S4b**. At 4 K, the spin polarization reaches up to 75%, corresponding to a majority spin percentage of 87%. In terms of TMR at room temperature, **Fig. S4c** displays ~6.4% at a bias current of 50 nA and ~5.5% at a bias current of 1 μA.

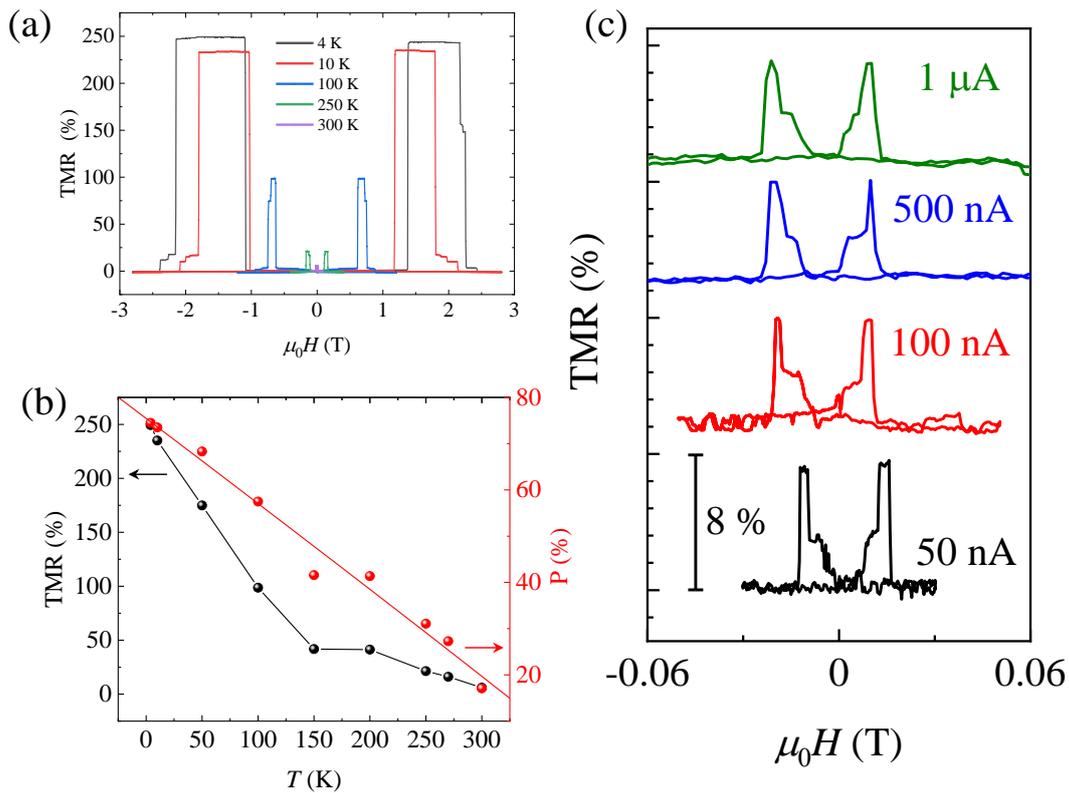

**Fig. S4. Fe$_3$GaTe$_2$ based magnetic tunnel junction. a,** TMR measured at various temperatures. **b,** Temperature dependence of TMR ratios (black) and spin polarization (red). **c,** TMR at 300 K under various bias currents.

**Supplemental Note 3: OTT effect under in-plane magnetic field.**

We further fabricated a WTe$_2$/Fe$_3$GeTe$_2$ device, where the OTT effect is also observed. Additionally, we demonstrate the OTT effect in varying in-plane magnetic fields. Notably, the negligible influence of the in-plane magnetic field on magnetization switching underscores that the switching torque predominantly arises from orbital moments, rather than spins.

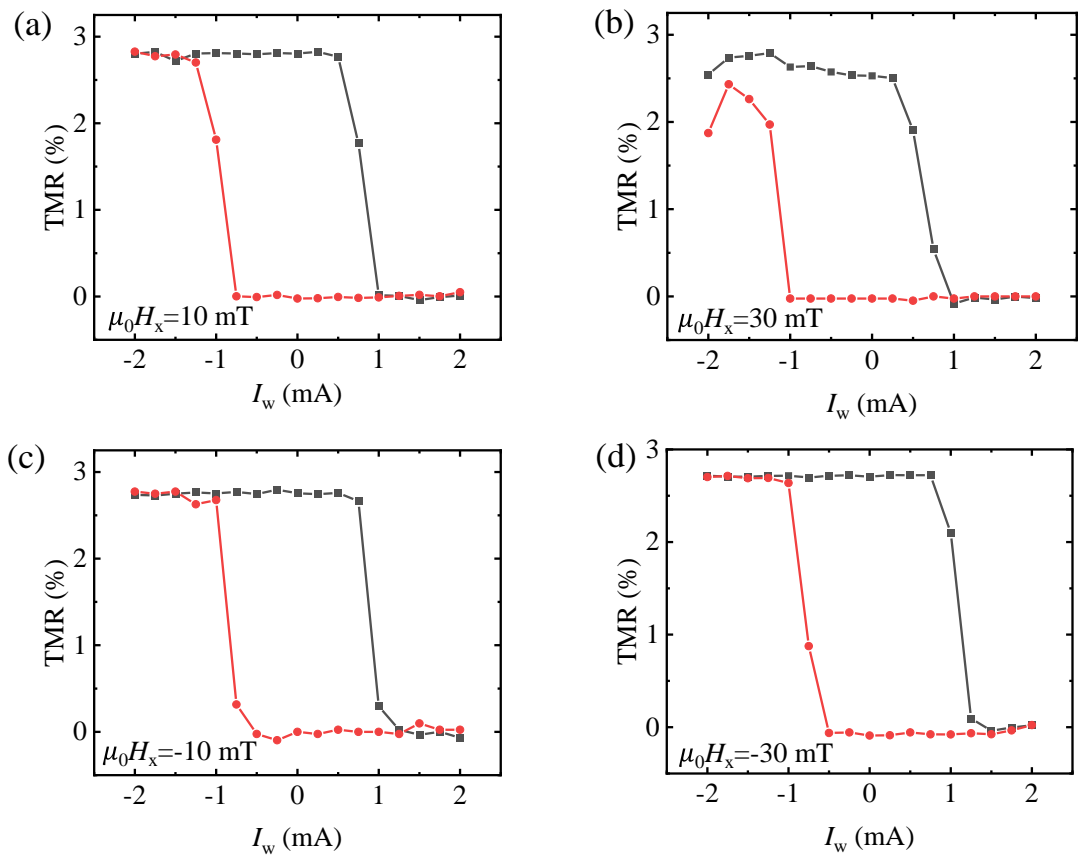

**Fig. S5. OTT-driven perpendicular magnetization switching in a WTe$_2$/Fe$_3$GeTe$_2$ device at 140 K with different in-plane magnetic fields.**

**Supplemental Note 4: Thermal effect in WTe$_2$/Fe$_3$GaTe$_2$ memory device.**

To exclude thermal effect as the main cause of magnetization switching, control experiments are carried out. Figure S6 shows the resistance of the MTJ measured when a 1.8 mA pulse current (larger than the critical magnetization switching current) was applied. The hysteresis behavior is clearly observed, indicating that the device temperature even under the large Joule heating is still below the Curie temperature of Fe$_3$GaTe$_2$.

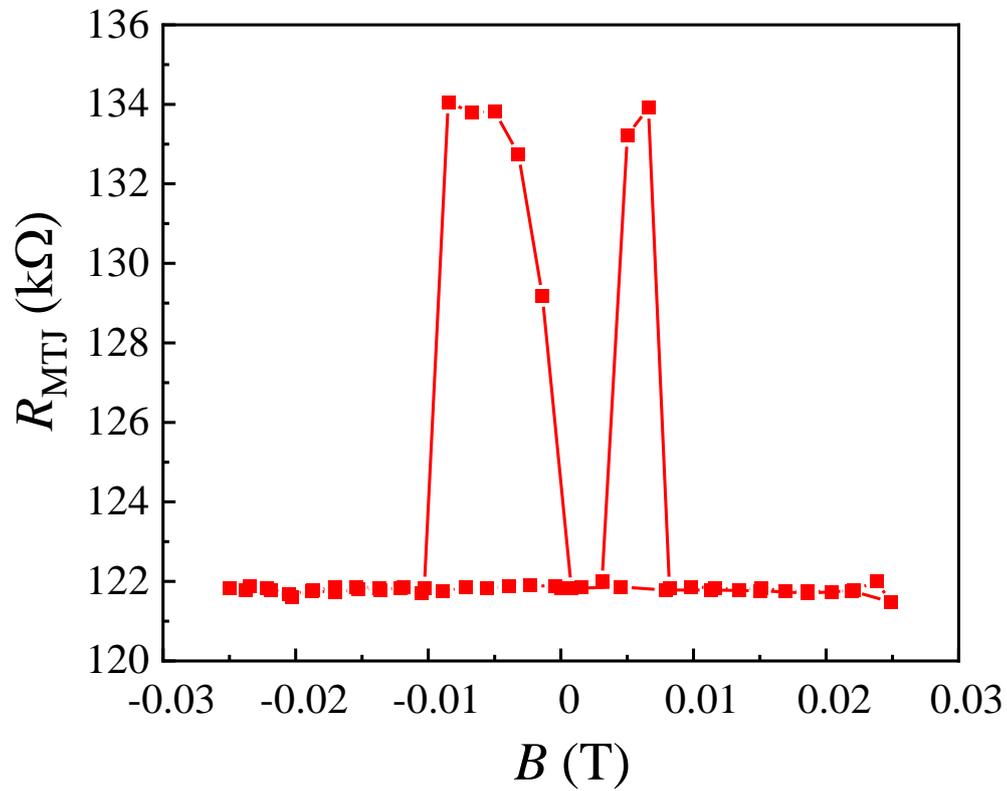

**Fig. S6. Resistance of the MTJ measured with $I_p$ = 1.8 mA at 300 K.**